\documentclass[prb,twocolumn]{revtex4}%

\usepackage{color}
\usepackage{amsmath}
\usepackage{graphicx}
\usepackage{textcomp}
\usepackage{tabularx}
\usepackage{ulem}

\newcommand{\be}{\begin{equation}}
\newcommand{\ee}{\end{equation}}
\newcommand{\bea}{\begin{eqnarray*}}
\newcommand{\eea}{\end{eqnarray*}}
\newcommand{\bean}{\begin{eqnarray}}
\newcommand{\eean}{\end{eqnarray}}
\def\lan{\langle}
\def\ran{\rangle}

\begin{document}

\draft
\title
{\bf Thermoelectric efficiency of quantum dot molecules at a high
temperature bias: the role of thermal-induced voltage}

\author{Chih-Chieh Chen$^{\dagger}$}
\address{Research Center for Applied Sciences, Academic Sinica,
Taipei, 11529 Taiwan}
\address{Department of Physics, University of Illinois at Urbana-Champaign, Urbana, Illinois 6180, USA}
\author{David M T Kuo$^{\dagger\dagger}$}
\address{Department of Electrical Engineering and Department of Physics, National Central
University, Chungli, 320 Taiwan}
\author{Yia-Chung Chang$^{*}$}
\address{Research Center for Applied Sciences, Academic Sinica,
Taipei, 11529 Taiwan}
\address{Department of Physics,
National Cheng Kung University, Tainan, 701 Taiwan}

\date{\today}

\begin{abstract}
The nonlinear electron and heat currents of quantum dot molecules
(QDMs) under a temperature bias are theoretically investigated,
including all correlation functions arising from electron Coulomb
interactions in QDMs. Unlike the case of double QDs, the maximum
efficiency of serially coupled triple QDs (SCTQD) occurs in the
orbital depletion regime owing to the interdot Coulomb blockade.
The electron current in SCTQD shows a bipolar oscillatory behavior
with respect to the variation of QD energy levels, whereas the
heat current does not show such a behavior. This is mainly
attributed to thermal-induced bias. In addition, we illustrate how
the efficiency of SCTQD is influenced by the external load
resistance, and phonon heat flow. Finally, a direction-dependent
electron current driven by a temperature bias has been
demonstrated for a SCTQD with staircase-like energy levels.
\end{abstract}

\maketitle
\section{Introduction}
Energy harvesting of heat dissipated from electronic circuits and
other heat sources is one of the most important energy issues.[1]
The realization of such type of energy harvesting typically relies
on the search of thermoelectric (TE) materials with high figure of
merits ($ZT$).[2] Impressive ZT values for quantum-dot superlattices
(QDSL) systems have been demonstrated experimentally.[3] The
enhancement of ZT mainly arises from the reduction of phonon thermal
conductivity in QDSL, which is due to the increased rate of phonon
scattering from the interface of quantum dots (QDs).[1,2] If the ZT
value can reach 3, the solid state cooler will have the potential to
replace conventional compressor-based air conditioners owing to its
long life time, low noise and low air pollution. Besides the search
of TE devices with large ZT value, the optimizing of nonlinear
thermoelectric behavior under high temperature bias is crucial for
the design of the next-generation energy harvesting engine
(EHE).[1,2]

Recently, a great deal of efforts was devoted to the studies of
the nonlinear response of thermoelectric devices under high
temperature bias. The nonlinear phonon flow of nanostructures with
respect to large temperature bias were investigated
experimentally[4] and theoretically.[5-8] The phonon thermal
rectification behavior of silicon nanowire (which has a very low
efficiency) was reported experimentally.[4] More recently, the
highly efficient electron thermal diode was reported in a
superconductor junction system.[9] However, such a thermal
rectification behavior only exists at very low temperatures.
Unlike heat rectifiers which are used to control the direction of
heat flow [4-9], the design of an {EHE} driven by a large
temperature bias needs to optimize the efficiency in the energy
transfer from the waste heat[1,2]. Although the energy harvesting
of coupled double QDs with size near $1\mu m$ was reported
experimentally and theoretically,[10-12] the large-size QDs
typically have charging energy ($U$) much larger than energy
separation ($\Delta E$ ). To design EHE operated at room
temperature, we need to consider QDs with nanoscale size,
satisfying the condition of $\Delta E/U \gg 1$ to emphasize the
focus of our current study.

So far, experimental studies of EHE made of nanoscale
semiconductor QD molecules (QDMs) or molecules have not been
reported, mainly due to technical difficulties[13] ( for example,
an isolated single nanowire or a molecular chain connected with
electrodes) and the {lack of good theoretical designs. Therefore,
it is desirable to have theoretical studies which can provide
useful guidelines for the advancement of nanoscale TE technology.
{The most challenging issue for theoretical studies arises from
the strong electron Coulomb interactions in semiconductor QDMs or
realistic molecules. Many techniques such as rate equation, master
equation and numerical renormalization group method were used to
examine the TE properties of QD junction system.[14-24]. Most
theoretical studies of TE properties have focused on the linear
response regime.[14-20] The TE properties of molecular junction
system beyond linear response were theoretically reported in
references[21-24]. However, the many-body effects arising from
orbital-filling on the nonlinear TE properties of QDMs or
molecules are still not clear. Such effects are crucial when the
energy levels of QDM are below the Fermi energy of electrodes.[20]
Under high temperature bias, the thermal-bias induced electrical
current will cause a voltage drop due to the load resistance,
which in turn will also modify the current. Thus, the theoretical
design of such an EHE must solve the thermal-induced electrical
voltage in a self-consistent way. In this article, we describe
such a self-consistent approach to study the nonlinear behavior of
EHE made of double QDs and serially coupled triple QDs (SCTQDs).
The comparison of the results for DQD and SCTQD allows one to see
the trend of increasing the length of chain of QD molecules. Our
studies are based on numerical method as described in [25], which
can suitably address the full many-body effect in the Coulomb
blockade regime for coupled multiple QDs. Due to the many-body
effect, the maximum efficiency of EHE made of the SCTQDs occurs
only in the orbital depletion regime. Meanwhile, we also clarify
how the efficiency of EHE is influenced by the physical parameters
of SCTQDs. In addition, we investigate an engine with
direction-dependent electrical output driven by a temperature-bias
for application as a novel TE devices.

\section{Theoretical method}
The insets of Figures 1 and 2 show the QD molecule (QDM) connected
to two metallic electrodes, one is in thermal contact with the
heat source at temperature $T_H$ (hot side) and the other with the
heat sink kept at temperature $T_C$ (cold side).  The heat flows
from the hot side through the QDM into the cold side. To reveal
the electron and heat currents driven by the temperature bias, we
consider the following Hamiltonian $H=H_0+H_{QD}$ for a QDM
junction system:

\begin{eqnarray}
H_0& = &\sum_{k,\sigma} \epsilon_k
a^{\dagger}_{k,\sigma}a_{k,\sigma}+ \sum_{k,\sigma} \epsilon_k
b^{\dagger}_{k,\sigma}b_{k,\sigma}\\ \nonumber &+&\sum_{k,\sigma}
V_{k,L}d^{\dagger}_{L,\sigma}a_{k,\sigma}
+\sum_{k,\sigma}V_{k,R}d^{\dagger}_{R,\sigma}b_{k,\sigma}+c.c
\end{eqnarray}
where the first two terms describe the free electron gas of left
and right electrodes (hot and cold sides).
$a^{\dagger}_{k,\sigma}$ ($b^{\dagger}_{k,\sigma}$) creates  an
electron of momentum $k$ and spin $\sigma$ with energy
$\epsilon_k$ in the left (right) electrode. $V_{k,\ell}$
($\ell=L,R$) describes the coupling between the electrodes and the
left (right) QD. $d^{\dagger}_{\ell,\sigma}$ ($d_{\ell,\sigma}$)
creates (destroys) an electron in the $\ell$-th dot.

\begin{small}
\begin{eqnarray}
H_{QD}&=& \sum_{\ell,\sigma} E_{\ell} n_{\ell,\sigma}+
\sum_{\ell} U_{\ell} n_{\ell,\sigma} n_{\ell,\bar\sigma}\\
\nonumber &+&\frac{1}{2}\sum_{\ell,j,\sigma,\sigma'}
U_{\ell,j}n_{\ell,\sigma}n_{j,\sigma'}
+\sum_{\ell,j,\sigma}t_{\ell,j} d^{\dagger}_{\ell,\sigma}
d_{j,\sigma},
\end{eqnarray}
\end{small}
where { $E_{\ell}$} is the spin-independent QD energy level, and
$n_{\ell,\sigma}=d^{\dagger}_{\ell,\sigma}d_{\ell,\sigma}$.
Notations $U_{\ell}$ and $U_{\ell,j}$ describe the intradot and
interdot Coulomb interactions, respectively. $t_{\ell,j}$
describes the electron interdot hopping. Noting that the interdot
Coulomb interactions as well as intradot Coulomb interactions play
a significant role on the electron transport in semiconductor QD
arrays or molecular chains.[15, 16] Because we are interested in
the case that the thermal energy is much smaller than intradot
Coulomb interactions, we consider QDs with only one energy level
per dot.

Using the Keldysh-Green's function technique,[26,27] the electron
and heat currents from reservoir $\alpha$ to the QDM junction are
calculated according to the Meir-Wingreen formula given by
\begin{eqnarray}
J_{\alpha}&=&\frac{ie}{\hbar}\sum_{j\sigma}\int
{\frac{d\epsilon}{2\pi}} \Gamma^\alpha_{j}(\epsilon) [
G^{<}_{j\sigma} (\epsilon)+ f_\alpha (\epsilon)(
G^{r}_{j\sigma}(\epsilon) \nonumber \\ &-&
G^{a}_{j\sigma}(\epsilon) ) ]\\ Q_{\alpha}
&=&\frac{i}{\hbar}\sum_{j\sigma}\int {\frac{d\epsilon}{2\pi}}
(\epsilon-\mu_{\alpha}) \Gamma^\alpha_{j}(\epsilon) [
G^{<}_{j\sigma} (\epsilon) + f_\alpha (\epsilon) \nonumber \\ & &(
G^{r}_{j\sigma}(\epsilon) - G^{a}_{j\sigma}(\epsilon) ) ],
\end{eqnarray}
Here $\Gamma^\alpha_{j}(\epsilon)=\sum_{k}
\delta(\epsilon-\epsilon_k)|V_{k,\alpha}|^2$ is the tunneling rate
between the left (right) reservoir and the left (right) QD of QDM.
For the simplicity, we consider the wide band limit of electrodes
to ignore energy-dependent tunneling rates
$\Gamma^{L(R)}_{L(R)}(\epsilon)=\Gamma_{L,(R)}$.
$f_{\alpha}(\epsilon)=1/\{\exp[(\epsilon-\mu_{\alpha})/k_BT_{\alpha}]+1\}$
denotes the Fermi distribution function for the $\alpha$-th
electrode, where $\mu_\alpha$  and $T_{\alpha}$ are the chemical
potential and the temperature of the $\alpha$ electrode.
$\mu_L-\mu_R=-e\Delta V$ and $T_L-T_R=\Delta T$. $e$, $\hbar$, and
$k_B$ denote the absolute value of electron charge, the Planck's
constant, and the Boltzmann constant, respectively.
$G^{<}_{j\sigma} (\epsilon)$, $G^{r}_{j\sigma}(\epsilon)$, and
$G^{a}_{j\sigma}(\epsilon)$ are the frequency domain
representations of the one-particle lesser, retarded, and advanced
Green's functions $G^{<}_{j\sigma}(t,t')=i\langle
d_{j,\sigma}^\dagger (t') d_{j,\sigma}(t) \rangle $,
$G^{r}_{j\sigma}(t,t')=-i\theta (t-t')\langle \{
d_{j,\sigma}(t),d_{j,\sigma}^\dagger (t') \} \rangle $, and
$G^{a}_{j\sigma}(t,t')=i\theta (t'-t)\langle \{
d_{j,\sigma}(t),d_{j,\sigma}^\dagger (t') \}  \rangle $,
respectively. These one-particle Green's functions are related
recursively to other Green's functions and correlation functions
via a hierarchy of equations of motion (EOM).[28] We truncate the
equation of motion by integrating out the leads degrees of freedom
using the Markov approximation, so the Kondo physics [29] is
ignored. By doing so we can focus on solving the Green's function
within the triple quantum dots system with the effects due to
coupling to leads approximated by a constant self-energy term, and
the hierarchy of EOM self-terminates at the level of $2N$-particle
Green's functions for the $N$-QD system (with $2N$ levels,
including spin). We have considered the Pauli exclusion principle
and charge conservation. For SCTQDs with one energy level in each
QD, there are $6$ energy levels (including spin). In the steady
state, the number of Green's functions (involving up to six
particles) described by $\langle d^\dagger_{i_1}\cdots
d^\dagger_{i_{n-1}} d_{j_1}...d_{j_n}(t) d^\dagger_{j}(t')\rangle$
($n=1,\cdots,6$) is given by $\sum_{n=1}^6 \binom{6}{n-1} \times
\binom{6}{n} \times 6=4752$. Using charge conservation ($U(1)$
symmetry), the number of correlation functions $\langle
d^\dagger_{i_1}...d^\dagger_{i_n} d_{j_1}...d_{j_n}\rangle$ needed
to be solved can be reduced to $\sum_{n=1}^6
(\binom{6}{n})^2=923$. The self-consistent solution to these
equations are solved numerically according to procedures described
in refs. [20] and [30] The algorithms employed are numerically
stable, and for equilibrium systems the code gives exactly the
same results as those obtained by exact diagonalization. In this
paper, we compare results obtained by the full calculation (method
A) and a simplified calculation (method B) to reveal the many-body
effect on the EHE efficiency. In method A, we calculate the
electron and heat currents of Eqs.~(3) and (4) by considering all
correlation functions resulting from electron Coulomb
interactions. Both methods are valid only in the Coulomb blockade
regime, not the Kondo regime. In the case of DQD the method A gets
exactly the same results for tunneling current as reported in
Ref.~[28].

With method A it is difficult to illustrate the behaviors of
electron and heat currents due to the lack of simple analytic
expressions. In method B, we neglect all correlations functions
except the two-electron correlation function for electrons on the
same site, whereas we still consider all many-body Green functions
(up to six electrons). This simplified method allows us to obtain
the expressions of electron and heat currents in terms of an
analytic function, ${\cal T}_{LR}(\epsilon)$ called transmission
coefficient. Their expressions are

\begin{eqnarray}
J=\frac{e}{h}\int d\epsilon {\cal T}_{LR}(\epsilon)
[f_L(\epsilon)-f_R(\epsilon)],
\end{eqnarray}
and
\begin{equation}
Q_{L/R}=\pm \frac{1}{h}\int d\epsilon ~(\epsilon-\mu_{L(R)}){\cal
T}_{LR}(\epsilon) [f_L(\epsilon)-f_R(\epsilon)].
\end{equation}
Because there are four possible states for each QD level (empty,
one spin-up electron, one spin-down electron, and two electrons),
${\cal T}_{LR}(\epsilon)$ contains $4^3=64$ configurations for the
SCTQD. The expression of ${\cal T}_{LR}(\epsilon)$ can be found in
Ref. 31, in which only one-particle occupation
numbers and two-particle on-site correlation functions used in the
Green's functions are considered. Method B requires much less
computation effort than the full calculation and can take
advantage of the analytic expression for ${\cal
T}_{LR}(\epsilon)$. Therefore, method B is very useful in
clarifying the physical mechanisms responsible for the results
obtained by the full calculation (method A).

To design an EHE driven by a high temperature-bias $\Delta T$, the
thermal induced voltage ($-eV_{th}=\mu_L-\mu_R$) across the
external load with conductance $G_{ext}=1/R_{ext}$ needs to be
calculated. To obtain $eV_{th}$, we have to solve
self-consistently all correlation functions appearing in the
electron current subject to the condition $G_{ext}V_{th}=J$, where
$J=(J_{L}+J_{R})/2$ is the net electron current. The heat current
satisfies the condition {$Q_L+Q_R=J\times V_{th}$}, which denotes
the work done by the EHE per unit time. The efficiency of EHE is
thus given by
\begin{equation}
\eta=|J*V_{th}|/Q_L.
\end{equation}
In the following discussions, we will illustrate the nonlinear
electron transport of EHE mostly based on the full calculation and some based on the simplified calculation (method B) for comparison.

\section{Results and discussion}

{An EHE made of a single QD with one energy level was
theoretically studied in Refs. \onlinecite{Nakp} and
\onlinecite{Leij}. To reveal the electron coherent tunneling
effect on the efficiency of EHE, the case of DQD under a fixed
temperature difference $\Delta T$ and electrical voltage $\Delta
V$ was studied in Ref. \onlinecite{Liu}. They then evaluate the
EHE efficiency according to $\eta=|J*V|/Q_L$. In the realistic
operation of an EHE, there is a fixed external load with
resistance $R_{ext}$. Since the voltage drop across the load must
satisfy the relation $\Delta V=J R_{ext}$, it can be argued that
in {\color{blue}Refs. 32-34} an external resistance
$R_{ext}=\Delta V/J$ was assumed. However, when the EHE efficiency
$\eta$ is examined as a function of some external parameter (such
as the gate voltage), keeping a fixed $\Delta V$ becomes
unphysical, since it implies a continuous change of the load
resistance as the external parameter varies. A more physical way
to study the dependence of $\eta$ is to calculate the
thermal-induced bias ($V_{th}$) (arising from the Seebeck effect)
self-consistently for a fixed load.

Here, we consider a fixed load with conductance
$G_{ext}=1/R_{ext}=0.2G_0$, where $G_0=2e^2_0/h$. Based on the
full calculation and Eq.~(7), we obtain the electron current
($J$), thermal-induced bias ($V_{th}$) and efficiency ($\eta$) of
DQD as a function of QD energy level tuned by gate-voltage
($E_L=E_R=E_F+30\Gamma_0-eV_g$) for various $T_C$ values with
$\Delta T=1\Gamma_0$. The results are shown in Fig. 1. We found
bipolar oscillatory behaviors for $J$ and $V_{th}$ with respect to
QD energy level, similar to the behavior of Seebeck coefficient
($S$). Such an oscillatory behavior was experimentally reported in
a single metallic QD case.[\onlinecite{Sven}] There are four main
structures in the $J$, $V_{th}$ and $\eta$ curves, which
correspond to processes of electrons tunneling through the DQD in
the one-, two-, three-, and four-electron states, respectively.
The maximum $\eta$ occurs either in the orbital-depletion regime
(with DQD in the one-electron state) or in the full
orbital-filling regime (with DQD occupied by four electrons). The
suppression of $\eta$ due to increasing $T_C$ is also illustrated
in this figure.

To gain deeper insight of the results shown in Fig. 1(c), we consider the expression of $\eta_{ZT}$ derived by
the classical approach given in Refs. 1 and 2.
\begin{equation}
\eta_{ZT}=(\frac{\Delta T}{T_C+\Delta T})\frac{m}{m+(1+m)^2/(ZT_H)+\bar T/T_H},
\end{equation}
where $m=G_{e}/G_{ext}$, and $Z=\frac{S^2 G_e}{\kappa}$. $G_e$,
$S=V_{th}/\Delta T$, and $\kappa=\kappa_e+\kappa_{ph}$ are the
electrical conductance, Seebeck coefficient, and thermal
conductance of the EHE. $\kappa_e$ and $\kappa_{ph}$ are the
electron and phonon thermal conductance, respectively. $\bar
T=(T_H+T_C)/2$. When $T_H$ approaches $T_C$, $ZT_H$ becomes the
dimensionless "Figure of merit" in the linear response regime.
Eq.~(8) reveals that the EHE becomes a Carnot engine with
$\eta_{C}=\Delta T/(T_C+\Delta T)$ as $ZT$ approaches infinity and
$m\gg 1$. Obviously, the suppression of maximum $\eta$ with
increasing $T_C$ can be illustrated by $\eta_{C}$. From Eq.~(8) we
can deduce that high-efficiency TE devices require large $\Delta T$
and $ZT$ values, which in general coincide with the condition of
small thermal conductivity.

Composite materials with high density of QDs embedded in a low
$\kappa_{ph}$ material can in general lower the thermal
conductivity.  For practical application, one should consider the
case of N-QDs between two metallic electrodes. However, due to the
very complicate many-body effect resulting from N-QDs, here we
only focus on the DQD and SCTQD systems and compare the results to
see the trend.

It is nontrivial to analyze the electron currents in the nonlinear
response regime with respect to a large temperature bias as many-body
effects can not be avoided in SCTQDs.[36-39] The analysis of
electron current spectra becomes very intriguing. To clarify how
the resonant channels of SCTQD resulting from electron Coulomb
interactions influence the electron transport, we compare the
calculated electron current for a fixed electrical voltage with
$e\Delta V=1\Gamma_0$ and $\Delta T=0$ to that for a short-circuit
case ($\Delta V=0$ or $R_{ext}=0$) with $k_B\Delta T=1\Gamma_0$,
and the results are shown in Fig. 2. In this case, it is
relatively easy to analyze the electron current spectra of SCTQDs.
Figure~2(a) shows the total occupation number ($N_t=\sum_{\sigma}
(\lan n_{L\sigma}\ran+\lan n_{C\sigma}\ran+\lan n_{R\sigma}\ran $
) of SCTQD without thermal bias ($k_B\Delta T=0$) as a function of
the applied gate voltage $V_g$ for three different temperatures
($k_BT_C=1, 3, 5\Gamma_0$). The electrical bias is set at $e\Delta
V=1\Gamma_0$. The staircase behavior of $N_t$ is due to the
charging effect arising from electron intradot and interdot
Coulomb interactions. The average occupancies in the center dot
($\lan n_{C\sigma}\ran=N_{C,\sigma}$) and outer dots ($\lan
n_{L\sigma}\ran (N_{L,\sigma}) =\lan n_{R\sigma}\ran
(N_{R,\sigma})$) are also plotted in Fig.~2(a) as dash-double-dots
and dash-dotted curves, respectively. Because of symmetry, the
average occupancies in two outer dots remain the same as $V_g$
varies, which leads to a jump of 2 for $N_t$ for the first two
steps.

The corresponding tunneling currents $J_{\Delta V}$ are plotted in
Fig.~2(b). The negative sign of $J_{\Delta V}$ labeled by
$\epsilon_1$ and $\epsilon_2$ indicates that the electron current
is flowing from the right electrode to the left electrode. The
tunneling currents are appreciable only in the regions where $N_t$
jumps a step, but become blocked when $N_t$ is flat as a function
of $eV_g$. $J_{max}$ is suppressed with increasing temperature
($T_C$). Although many efforts have been devoted to studies of
electron transport through SCTQDs under an applied
bias [36-39], not many literatures studied the
electron current through SCTQD under high temperature bias.

Fig. 2(c) shows the electron current driven by a temperature bias
for various values of $k_BT_C$ with $\Delta V=0$. {We note that
$J_{\Delta V}$ and $J_{\Delta T}$ are vanishingly small when $N_t$
varies from four to five. This is due to the lack of resonant
channel in SCTQD, which leads to
$E_L+U_0+U_{LC}=E_R+U_0+U_{CR}\neq E_C+2U_{LC}+2U_{CR}$ and
therefore the electron transport is blockaded. Such an effect also
exists for the change of $N_t$ from five electrons to six
electrons (not shown here).

%{\color{blue} Note, in the case of triangular TQD [25] single hole
%(corresponding to $N_t$ from five electrons to six electrons)
%shows a very good transport behavior. Therefore, the single-hole
%transport blockaded in Fig. 2 arises from the structure effect.}}

%In Fig. 1 the applied bias ($e\Delta V=-\Gamma_0$) is irrelevant with $\Delta T$.
In the practical operation of EHE, a temperature bias $\Delta T$
should induce a thermal voltage $V_{th}$  ($
-eV_{th}=\mu_L-\mu_R$) which depends not only on the load
conductance $G_{ext}$ but also on the correlation functions
resulting from electron Coulomb interactions. Such behavior is
illustrated in Fig. 3, which shows the electron current ($J$),
thermal-bias induced voltage ($V_{th}$) and $\eta$ driven by a
temperature bias at $k_B\Delta T=1\Gamma_0$ for various values of
$k_BT_C$. Comparing with Fig.~2(c), we see that the behavior of
electron current ($J$) is qualitatively similar to the case with
$R_{ext}=0$. However, the magnitude is reduced by about 30\% when
$G_{ext}=0.2G_0$ due to the counter balancing effect through the
thermal-bias induced voltage $V_{th}$. Both $J$ and $-eV_{th}$
shown in Fig.~3  display a bipolar oscillatory behavior, similar
to that shown in  Fig.~1(a). In the orbital depletion regime, the
behaviors of SCTQD are very similar to those of DQD, whereas
deviation occurs (with the efficiency ($\eta$) lowered by about
30\% compared with DQD) when QD energy levels are below $E_F$.
$\eta$ becomes even lower at higher $V_g$ due to the lack of
resonant channels in SCTQD (caused by the blockade of electron
transport from interdot Coulomb interactions). In the absence of
$U_I$, $\eta$ can be higher even for higher $V_g$. We see that the
highest efficiency of EHE occurs near the transition where $N_t$
goes from 0 to 1 (with $eV_g\approx 25 \Gamma_0$), which is in the
orbital-depletion regime.

To reveal the importance of electron correlation arising from many
body effect, the physical quantities of Fig.~3 are recalculated by
method B. The resulting curves are shown in Fig.~4, which have
one-to-one correspondence to those of Fig.~3. For the low-filling
situation (with $eV_g < 30 \Gamma_0$), the results agree very well
with the full-calculation results shown in Fig.~3. On the other
hand, there are appreciable differences between the two results as
$N_t$ exceeds 1 (with $eV_g > 30\Gamma_0$), although their behaviors
are qualitatively the same for $eV_g$ up $100\Gamma_0$. This implies
that a simplified model without considering interdot correlation
functions is sufficient to model the main characteristics of the EHE
made of SCTQDs in the low-filling regime ($N_t\le 1$).

So far, we have fixed $G_{ext}=0.2 G_0$ and neglected the phonon
heat flow ($Q_{ph}=0$). In the inset of Fig. 4, we show results
for $\eta=|J\times V_{th}|/(Q_L+Q_{ph})$ versus $V_g$ for four
different values of $G_{ext}$ at $k_BT_c=1\Gamma_0$ with
$k_B\Delta T=1\Gamma_0$ and $t_C=1\Gamma_0$, where we have
included the effect of phonon heat flow given by a simple model
$Q_{ph}=\kappa_{ph,0}F_s \Delta T$. ($F_s$ is the correction
factor describing the phonon scattering resulting from the surface
of nanowires and the QDs) Here,
$\kappa_{ph,0}=\frac{\pi^2k^2_B\bar T}{3h}$ is the universal
phonon thermal conductance arising from acoustic phonon
confinement in a nanowire. It is generally accepted that the
linear term of phonon thermal conductivity $\kappa_{ph,0}$ can
well illustrate the behavior of silicon nanowire even at room
temperature.[40] In a nanowire filled with QDs considered in our
paper, photon scattering is dominated by the defect scattering
(due to structure difference between QDs and nanowire), which
implies that the phonon mean-free path, $\ell$ is not sensitive to
the temperature variation. The linear dependence in temperature
comes from the specific heat $C$, which is proportional the phonon
density of states in one dimension, thus linearly proportional to
temperature. According to the relation $\kappa=Cv\ell$, we obtain
a linear temperature dependence for $\kappa$, since the sound
velocity ($v$) is also not very sensitive to temperature.
Furthermore, the physics for the reduction of $\eta$ in the
presence of $Q_{ph}$ will remain qualitatively similar even if
there is small nonlinear effect in $\kappa_{ph}$ in the
temperature range considered. Recently, phonon thermal
conductivity in the Kondo regime (extremely low temperature
regime) has been investigated in Ref. \onlinecite{Zhang}. The
maximum efficiency is obtained at $G_{ext}=0.05G_0$. Meanwhile,
the maximum $\eta$ for $G_{ext}=0.2G_0$ (blue line) is around 0.08
including the effect of $Q_{ph}$, which is about one half of the
value obtained with $Q_{ph}=0$ as shown by the black solid line in
Fig. 4(c). According to Eq.~(6), the optimized $\eta_{ZT}$ occurs
at $G^{o}_{ext}=G_e/\sqrt{1+Z\bar T}$. Thus, $\eta_{max}$ will
occur at vanishingly small $G_{ext}$ if $Z\bar T$ becomes very
large. When $Z\bar T=2$, $G^0_{ext}=G_e/\sqrt{3}\approx 0.047G_0$
for $G_e=0.08G_0$. This is consistent with the results shown in
the inset of Fig. 4, where $\eta_{max}$ occurs near
$G^0_{ext}=0.05 G_0$.

To illustrate the effect of thermal-bias induced voltage, we also
calculate the electron current and heat current of SCTQD as
functions of gate voltage $V_g$ for the case of a fixed electrical
voltage with $e\Delta V=1 \Gamma_0$  based on method B. Note that
this situation corresponds to a load conductance $G_{ext}$ which
varies with $V_g$ with the relation $G_{ext}=J/\Delta V$. Fig.
5(a) shows the bipolar behavior of the electron current, which is
qualitatively similar to the results shown in Fig.~4(a) except
that the magnitude of $J$ is much smaller here. In Fig~5(b) we
show the heat currents obtained for fixed $e\Delta V=1 \Gamma_0$
(solid curves) as well as the results for a fixed load with
$G_{ext}=0.2G_0$ (dashed curves). It is noted that $Q_L$ becomes
negative for $k_BT_C=3 \Gamma_0$ (red solid curve) and $5
\Gamma_0$ (blue solid curve) for $eV_g$ around $28 \Gamma_0$. Such
negative values are caused by the negative load conductance
$G_{ext}=J/\Delta V$ implicitly adopted, where $J$ becomes
negative for $\Delta V=1\Gamma_0$. Such a situation does not
correspond to a realistic operation of EHE. Once we consider a
fixed load and find the self-consistent solution to $V_{th}$, the
heat currents are always positive (as shown by dashed curves).
Because of this issue, in {\color{blue}Refs. 32-34}, the $\eta$
value of EHE can only be evaluated for the area of $Q_L$ larger
than zero. From the comparison of results in Fig. 5(b), we see
that it is important to include $V_{th}$ in a self-consistent way
in the optimization of $\eta$ for EHE. The inset of Fig. 5 shows
the efficiency of EHE as functions of $G_{ext}$ for different
$k_B\Delta T $ values at $ eV_g=28\Gamma_0$ and $k_BT_C=1\Gamma_0$
in the absence of $Q_{ph}$. The maximum $\eta$ occurs at
$G_{ext}~0.01G_0$, which is smaller than that for the case with
$Q_{ph}$. This also indicates that the $ZT$ value of SCTQD is
highly enhanced in the absence of $\kappa_{ph}$. For a fixed
$G_{ext}$, $\eta$ is enhanced with increasing $\Delta T$ in the
orbital depletion region.

So far, we have focused on a fixed electron hopping strength
$t_C=1\Gamma_0$. It is also interesting to examine how $\eta$ is
influenced when $t_c$ increases. We plot $J$, $Q_L$ and $\eta$ as
functions of $t_C$ for $E_0=E_F+2\Gamma_0$ and $eV_g=0$ in Fig. 6
based on method B. To illustrate the results of Fig. 6, the
approximated expression of ${\cal T}^1_{LR}(\epsilon)$ is given
below:
\begin{equation}
{\cal T}^{1}_{LR}(\epsilon) =\frac{4\Gamma_L\Gamma_R
P_{1}t^2_{LC}t^2_{CR}}
{|\mu_1\mu_2\mu_3-t^2_{CR}\mu_1-t^2_{LC}\mu_3|^2},
\end{equation}
where $\mu_1=\epsilon-E_L+i\Gamma_L$, {$\mu_2=\epsilon-E_C$} and
$\mu_3=\epsilon-E_R+i\Gamma_R$.
$P_{1}=(1-N_{L,\bar\sigma})(1-N_{C,\bar\sigma}-N_{C,\sigma}+c_C)
(1-N_{R,\bar\sigma}-N_{R,\sigma}+c_R)$ denotes the probability
weight of electron transport through SCTQD in an empty state,
which is determined by the one particle occupation number
($N_{\ell,\bar\sigma}$) and on-site two particle correlation
functions ($c_{\ell}$) resulting from electron Coulomb
interactions. Only, the first of 64 configurations for
SCTQD is included in Eq. (9), because the QD energy levels
are above $E_F$. $P_1$ equals to one in the absence of electron
Coulomb interactions.[30] We drive the expression of tunneling
current in the small tunneling rate limit ($\Gamma/k_B \bar T \ll 1$)
under the assumption $\Delta T/{\bar T} \ll 1$ and obtain

\begin{equation}
J=\frac{2e}{h}\frac{\pi \Gamma P_1}{k_B\bar T^2} \frac{4
t^2_{LC}t^2_{CR}(E_0-E_F)}{(t^2_{LC}+t^2_{CR}+\Gamma^2)^2}
\frac{\Delta T} {cosh^2\frac{E_0-E_F}{2k_B\bar T}}.
\end{equation}
From Eq.~(10), we see that the maximum $J$ occurs when
$t_{LC}=t_{CR}=t_C$. Thus, non-uniform electron hopping strength
tends to reduce $J$. Meanwhile, the maximum $J$ and $\eta$ occur
at $t_c=\Gamma_0/\sqrt{2}$, which well explains the results of
Fig. 6(a). In addition, the suppression of $J$ with increasing
$T_C$ can also be described by Eq. (10). Note that tunneling
current arising from $V_{th}$ has been neglected in Eq. (10). From
the results of Fig. 6, we see that $J$ and $Q_L$ are not a
monotonic function of $t_C$. Because $t_{LR}=0$ in this
calculation, we do not observe the interesting quantum
interference effect (QIE). In Ref. [20], how QIE influences
electrical conductance, Seebeck coefficient and electron thermal
conductance was discussed in the case of triangular TQD.

To further examine the behavior of the EHE efficiency, we plot in
Fig.~7 $J$, $Q_L$ and $\eta$ as functions of $V_g$ for various
values of $k_B\Delta T$ with $T_C$ fixed at $1\Gamma_0$ and
$G_{ext}=0.05G_0$. We see that the peak values of $J$, $Q_L$ and
$\eta$ all increase with $\Delta T$. It is worth noting that $Q_L$
is positive in the entire parameter space (unlike the electron
current which shows bipolar oscillatory behavior with respect to
QD energy level). The results of Fig.~7 indicate that a high
efficiency engine with large electrical outputs requires a high
temperature bias, which exists only in a system with high thermal
resistivity (phonon glass). Serially coupled QDs can enhance the
phonon scattering and thus reduce thermal conductivity. Therefore,
a long chain of QD molecules is desirable for implementing EHE
with high efficiency. As for the optimization length of QD
molecules, this problem is beyond the scope of present article.
There are two reasons: (a) the model of phonon heat flow is too
simple to fully catch the realistic phonon heat flow magnitude,
and (b) the calculation of QD number beyond three requires the
high cost computing time. For most conventional TE devices, there
exist a trade-off between high efficiency and large output power.
Based on the results presented in Figs.~(3)-(6), it is concluded
that a high-efficiency EHE should operate in the low-filling
regime. Because the effect of $Q_{ph}$ is important (as eluded in
the inset of Fig.~4), we also calculate the EHE efficiency
including the $Q_{ph}$ effect. The results for $k_B\Delta
T=3\Gamma_0$ are shown as triangles, which is to be compared with
the dotted line of Fig. 7(c), Obviously, $\eta_{max}$ is
suppressed when $Q_{ph}$ is included. However, we note that the
maximum $\eta$ can still reach a maximum close to 0.2 in the
presence of $Q_{ph}$. This is considered high efficiency when
compared with conventional heat engines[1,2].

When there is size/shape variation in serially coupled QDs, three
important physical quantities including the electron Coulomb
interactions, tunneling rates and QD energy levels will be
changed. In the depletion regime with the best engine efficiency,
the energy level fluctuation (ELF) of QDs will cause a significant
effect on the ZT values. Therefore, it is desirable to examine the
ELF effect on the electron current (or EHE efficiency) of SCTQD.
In general, we found that the EHE efficiency is suppressed by ELF
in SCTQDs mainly due to the reduced electron current. However, we
found that a stair-case alignment of QD energy levels can be used
to design an engine with direction-dependent electrical output. In
Fig.~8, the electron current ($J$) and thermal voltage ($V_{th}$)
are calculated for an SCTQD with staircase-like alignment of
energy levels: $E_L=E_R+2\Delta$, $E_C=E_R+\Delta$ and
$E_R=E_F+10\Gamma_0$, where $\Delta$ is the QD energy level
difference. In Figs (1)-(7), we have neglected the voltage drop
across the dots due to $V_{th}$. To examine such an effect we show
in Fig. 8 the influence due to the change of outer QD energy
levels arising from $V_{th}$, which follows the relation
$\epsilon_{L(R)}=E_{L(R)}\mp D_\eta eV_{th}$. It's worth noting
that the tunable factor $D_{\eta}$ is mainly determined by the QD
separation. Here, we adopt $D_{\eta}=0.3$. For $\Delta T > 0
(\Delta T < 0)$, $T_H$ is on the left (right) electrode(See insets
of Fig.~8(a)). The forward (backward) currents ($J_{F(B)}$) are
positive (negative), while $V_{th}$ has opposite sign with respect
to $J$. Both forward and backward
 electron currents have a nonlinear dependence on
$\Delta T$. With increasing $\Delta$, the electron currents (or
electrical powers) are suppressed in the wide temperature bias
regime. For $\Delta=0$ (solid black curve), the electron current
shows no directionality, while for $\Delta=2\Gamma_0$ the
direction-dependent electron current becomes apparent. This
directionality of electron current can be qualitatively explained
as follows. When $\Delta T > 0$, $\epsilon_{L}$ and $\epsilon_R$
become aligned with $E_C$ as $-eV_{th}$ changes to around
$-2\Gamma_0$, while for $\Delta T < 0$, $\epsilon_L$ and
$\epsilon_R$ are tuned further away from $E_C$. Therefore, QD
energy level shift due to the thermal-induced voltage can play a
remarkable role for the current rectification effect in SCTQD with
staircase-like energy levels. The energy level shift of QDs
arising from thermal voltage was experimentally observed in the
DQD realized by lithographic technique.\onlinecite{Thier} The
experiment of Ref. \onlinecite{Thier} is limited to the low
temperature regime with a small temperature bias, since the QDs
considered are large and the charging energies are much larger
than the energy level separation. If $V_{th}$ is turned off, we
can no longer observe direction-dependent tunneling current under
temperature bias even though SCTQD has site-dependent QD energy
levels.

Let's define the electron current rectification efficiency as
$\eta_R=(J_F-|J_B|)|/(J_F+|J_B|)$, which is irrelevant to heat
flows. The calculated $\eta_R$ as a function of temperature bias
under various conditions is shown in Fig.~9. Figure~9(a) shows
$\eta_R$ for various values of $\Delta$ with $t_C=3\Gamma_0$. We
see that the highest rectification occurs when $\Delta=2\Gamma_0$
with $\eta_R$ approaching 0.2 at the high $\Delta T$ limit. The
rectification efficiency actually becomes poorer if $\Delta$ is
too large. Unlike the case with $\Delta=2\Gamma_0$, $\eta_R$
decreases with increasing $\Delta T$ for $\Delta=4$ and
$6\Gamma_0$. To reveal the electron correlation effects, we also
calculate $\eta_R$ with method B and plot the corresponding curves
with triangle marks in Fig.~9(a). It is found that the
rectification efficiency is overestimated in this simplified
model. When a temperature bias increases significantly, the total
occupation ($N_t$) increases. Therefore, electron-correlation
effect becomes strong. In particular, the interdot two-electron
correlation functions can no longer be ignored. This explains why
the numerical results of method B ( with less correlation
functions) becomes overestimated. Figure~9(b) shows $\eta_R$ for
$\Delta=2\Gamma_0$ for different electron hopping strengths
($t_C=0.5, 1,$ and $2 \Gamma_0$). $\eta_R$ is found to be largest
for $t_c=2\Gamma_0$ (dotted line), which is also larger than that
for $t_c=3\Gamma_0$ as shown in Fig.~9(a). Thus, the electron
current rectification efficiency is not a monotonic function of
$t_c$. In Fig.~9(c), we consider the effect of varying the
temperature of the cold side, $T_C$. The results indicate that the
maximum $\eta_R$ reduces with increasing $T_C$.
%The behaviors of Fig. 8(b) and 8(c) can be related  to the behaviors of thermal-induced voltages $V_{th}$ (not shown)}.

Nonlinear thermoelectric effects of nanostructures for developing
new applications have been reviewed in a recent
article.[\onlinecite{Zhu}] For phonon rectifiers, it is very
difficult to realize "phonontronics" due to large leakage of
phonon flow arising from acoustic phonons, which are difficult to
confine.[4-6,12] The heat rectification phenomena of electrons can
only exist at low temperatures, where phonon flows can be
suppressed.[7-9,12] On the other hand, the electron current
rectification shown in Fig.~9 will be unaffected by
the phonon flow. Therefore an EHE made of serially coupled QDs
with direction-dependent electrical current may prove useful in
the advancement of nonlinear thermoelectric
devices.[\onlinecite{Soth}]

\section{Summary}
The electron/heat transport in nanoscale semiconductor QDM driven
by a finite temperature bias is theoretically studied for the
application of EHE, which converts thermal energy into electrical
power. Our studies illustrate that the efficiency of the EHE made
of QDM must be evaluated by  Eqs. (3) and (4) under the constraint
$G_{ext}~V_{th}=J$, in which $V_{th}$ induced by $\Delta T$ should
be calculated self-consistently, otherwise the $J$ and $Q$ will
show nonphysical features. We have demonstrated that an EHE made
of a DQD has high efficiency  either in the charge-depletion or
full-filling regime, whereas an EHE made of SCTQD  prefers the
charge-depletion regime due to the lack of resonant channels in
the full-filling regime. We found that the EHE performance is
degraded by the energy-level fluctuation (ELF) in QDs, which may
arise from QD size variation or energy level shift caused by
thermal voltage $V_{th}$. $\eta_{max}$ of EHE is seriously
suppressed in the presence of phonon thermal conductance. QDMs
have promising potential for realizing high-efficiency EHEs due to
their low phonon thermal conductance. The direction-dependent
electron current is illustrated by an SCTQD with staircase-like
energy levels. The thermal voltage yielded by temperature bias
plays a remarkable role in the design of an engine with
bidirectional current driven by a temperature bias. The results of
Fig.~9(a) clearly reveal that interdot electron correlation
functions arising from electron Coulomb interactions (considered
in method A) play a significant role in the high temperature bias
regime. The condition of $\Delta E/U \gg 1$ for each QD can be
satisfied for small molecules such as benzene. (See Refs. [15] and
[16]). Therefore, our study is applicable for studying finite
benzene chain.

%\begin{flushleft}

%\end{flushleft}

\begin{flushleft}
{\bf Acknowledgments}\\
\end{flushleft}
This work was supported by the National Science Council of the
Republic of China under Contract Nos. MOST 103-2112-M-008-009-MY3 and MOST 104-2112-M-001-009-MY2.

\mbox{}\\
${}^{\dagger}$Present address: Department of Physics, Zhejiang University, Hangzhou 310027, China\\
${}^{\dagger\dagger}$E-mail address: mtkuo@ee.ncu.edu.tw\\
${}^*$E-mail address: yiachang@gate.sinica.edu.tw\\

\setcounter{section}{0}

\renewcommand{\theequation}{\mbox{A.\arabic{equation}}} %\section{Appendix}
\setcounter{equation}{0} % reset counter

%\section{}
%\subsection{Derivation of the tunneling current formula using Dyson's equations\label{App:TC_l} }
\mbox{}\\
%{\bf Appendix A. }

\newpage
%\clearpage
%\section*{Figure captions}
\clearpage
\begin{figure}[h]
\centering
\includegraphics[angle=-90,scale=0.3]{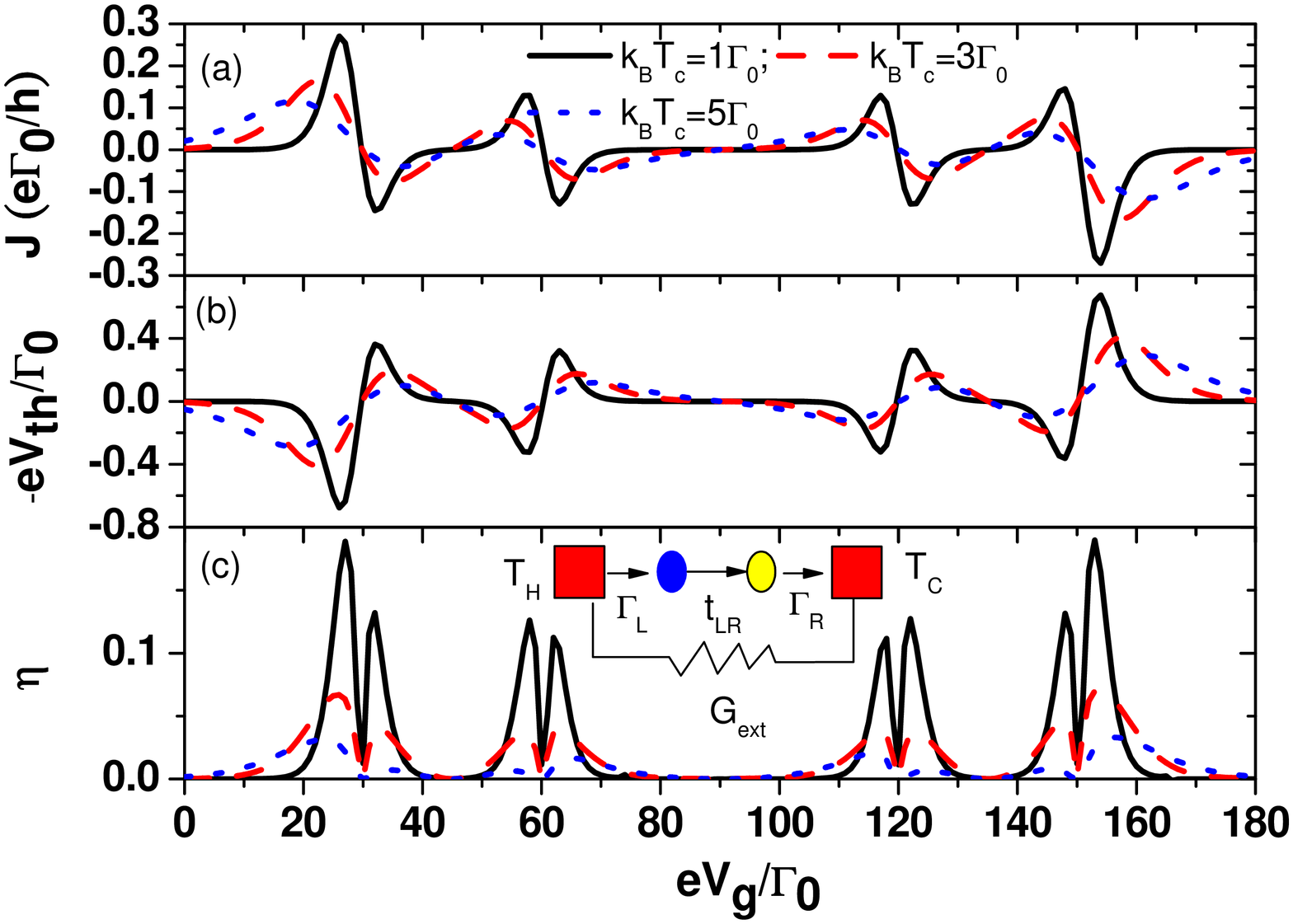}
\caption{Full many-body calculation results for (a) Electron
current ($J$), (b) thermal voltage ($eV_{th}/\Gamma_0$) and (c)
efficiency ($\eta$) of DQD as functions of gate voltage $V_g$
($E_{\ell}=E_0=E_F+30\Gamma_0-eV_g$) for various values of  $T_C$.
To find the ratio of $\eta$ to the Carnot efficiency, we should
multipy $\eta$ in (c) by a scaling factor $(T_C+\Delta T)/\Delta
T$, which is 2, 4, and 6 for $k_B T_C=1, 3$, and  $5
\Gamma_0$,respectively.  We have used the following physical
parameters $t_{LR}=1\Gamma_0$, $U_{\ell}=60\Gamma_0$,
$U_{LR}=30\Gamma_0$, and $\Gamma_L=\Gamma_R=\Gamma=1\Gamma_0$.
$G_{ext}$ is set to $0.2G_0$. }
\end{figure}

\begin{figure}[h]
\centering
\includegraphics[angle=-90,scale=0.3]{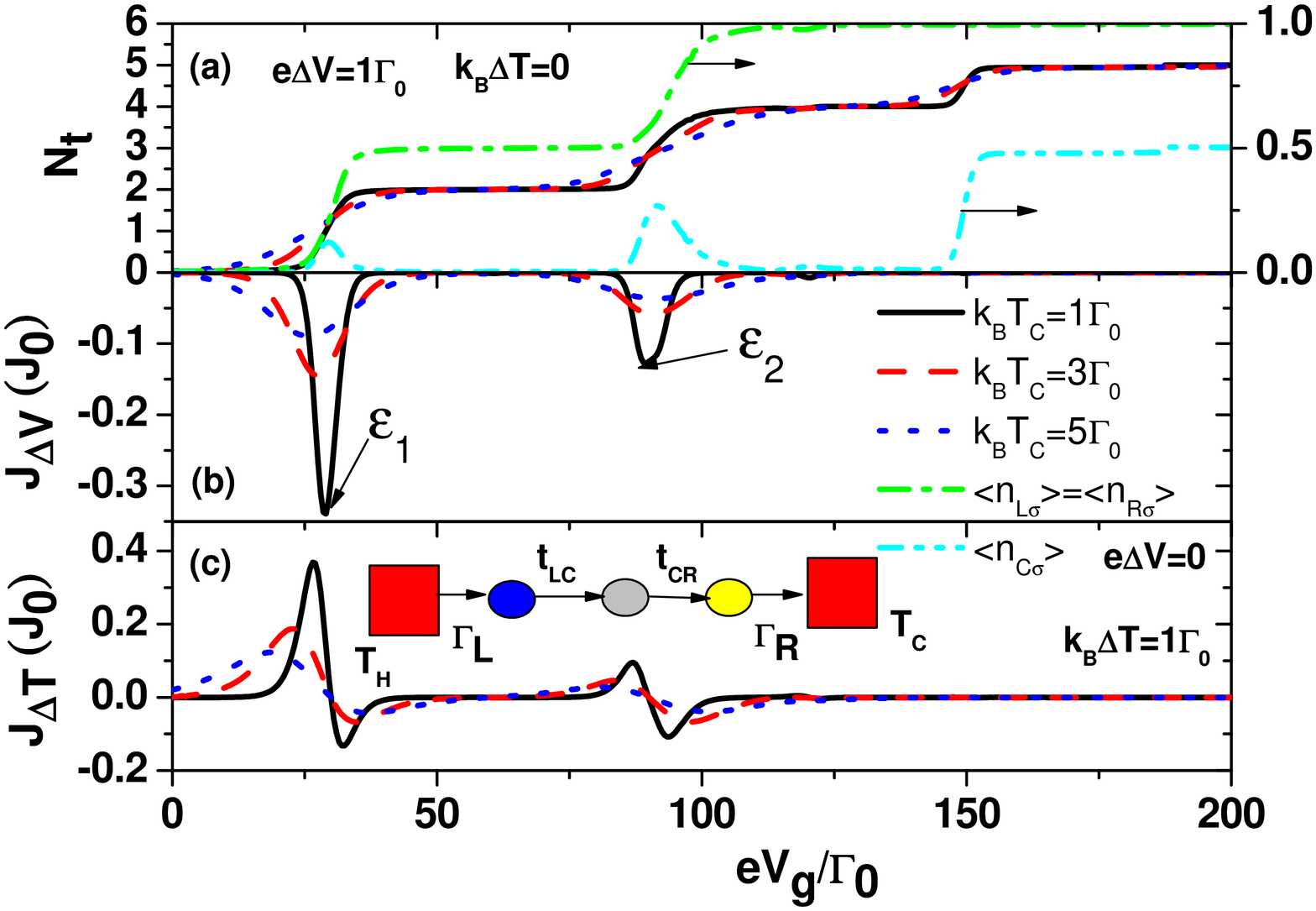}
\caption{Full many-body calculation results of SCTQD for (a) Total
occupation number, (b) electron current ($J_{\Delta V}$) due to
fixed electrical bias $e\Delta V=1\Gamma_0$ and $k_B\Delta T=0$ ,
and (c) electron current ($J_{\Delta T}$) due to fixed temperature
bias $\Delta T$ as functions of the gate voltage $V_g$
($E_{\ell}=E_0=E_F+30\Gamma_0-eV_g$) for various values of $T_C$
with $k_B\Delta T=1\Gamma_0$ and $e\Delta V=0$.  We used the
following physical parameters $t_{LC}=t_{CR}=1\Gamma_0$,
$t_{LR}=0$, $U_{\ell}=60\Gamma_0$, $U_{LC}=U_{CR}=30\Gamma_0$, and
$\Gamma_L=\Gamma_R=\Gamma=1\Gamma_0$. $J_{0}=e\Gamma_0/h$. Both
$R_{ext}$ and $Q_{ph}$ are set to zero.}
\end{figure}

\begin{figure}[h]
\centering
\includegraphics[angle=-90,scale=0.3]{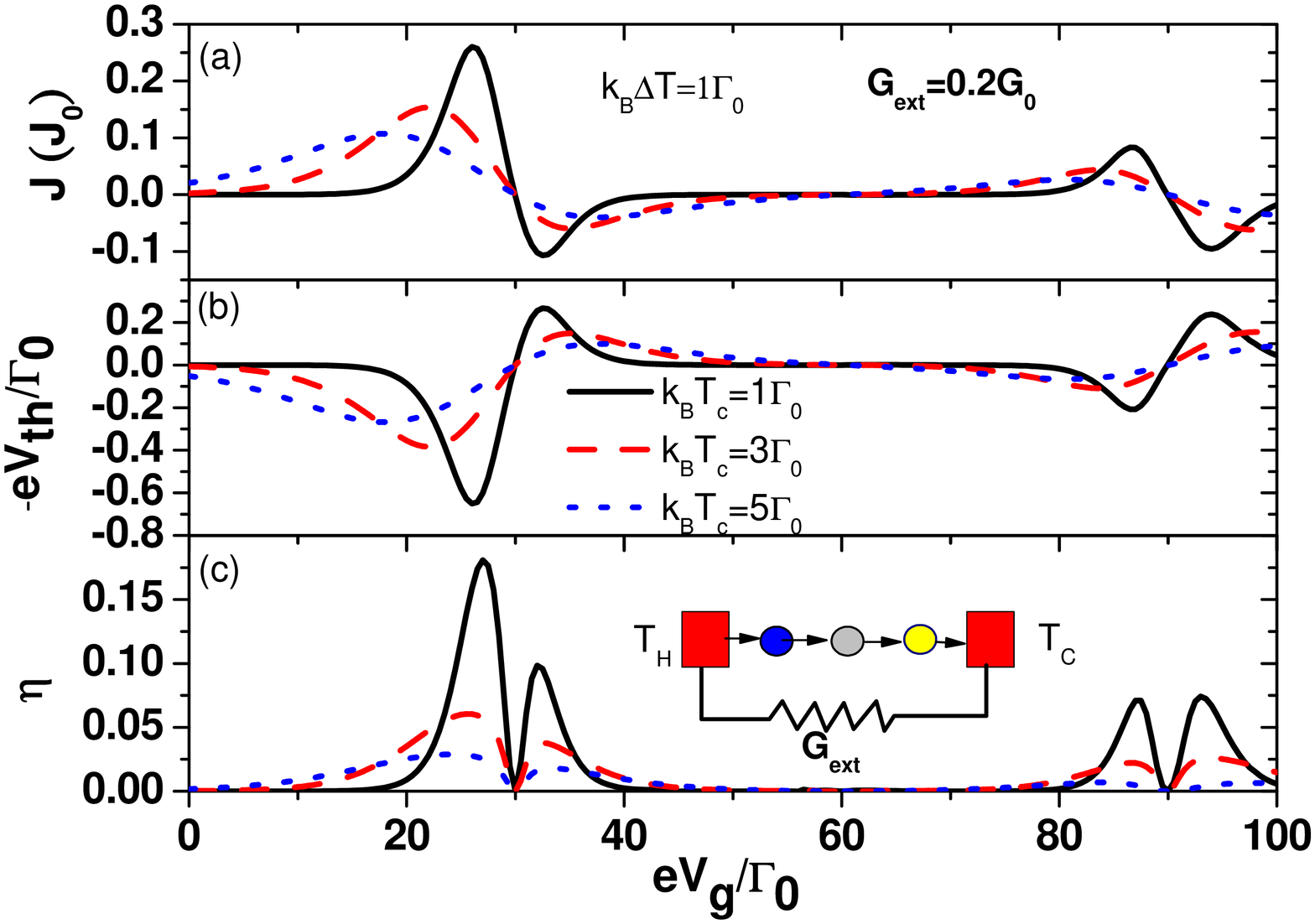}
\caption{Full  many-body calculation results for (a){Electron}
current ($J$), (b) thermal voltage ($eV_{th}$) and (c) EHE
efficiency ($\eta$) of SCTQD as functions of gate voltage $V_g$
($E_{\ell}=E_F+30\Gamma_0-eV_g$) in SCTQD for  various values of
$T_C$ with $k_B\Delta T=1\Gamma_0$. Other physical parameters are
the same as those of Fig. 2. $G_{ext}=0.2 G_0$ and $Q_{ph}=0$.}
\end{figure}

\begin{figure}[h]
\centering
\includegraphics[angle=-90,scale=0.3]{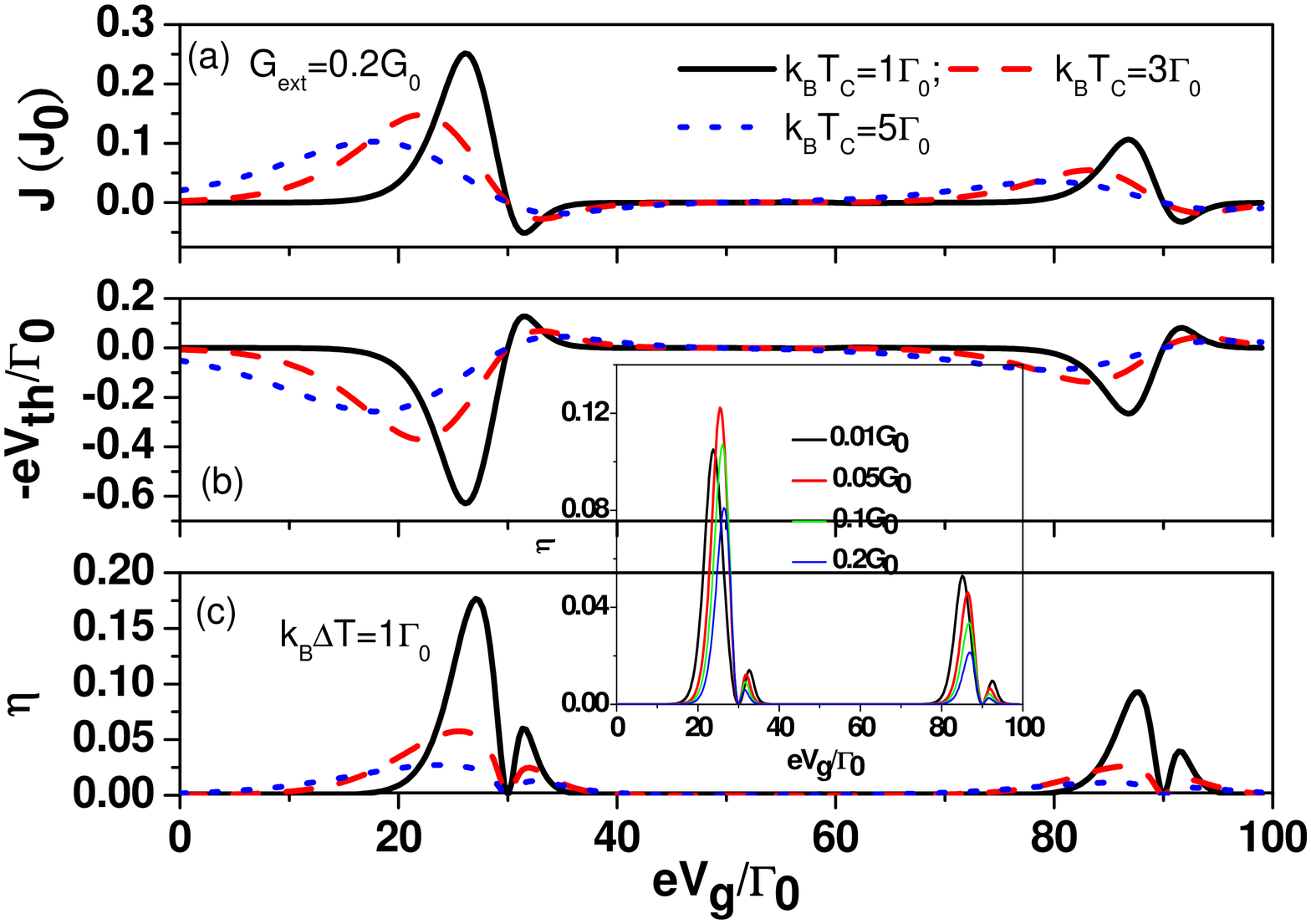}
\caption{Results obtained by method B for (a){Electron current} ($J$), (b) thermal voltage
($eV_{th}$) and (c) EHE efficiency ($\eta$) of SCTQD as functions of gate voltage $V_g$  ($E_{\ell}=E_F+30\Gamma_0-eV_g$) for various values of $T_C$ with $k_B\Delta T=1\Gamma_0$, $G_{ext}=0.2 G_0$, and $Q_{ph}=0$. The inset of Fig. 4 shows the
$\eta$ including the effect of $Q_{ph}$  for four $G_{ext}$ values; $0.01,0.05,0.1$ and $0.2G_0$ at
$k_BT_C=1\Gamma_0$.}
\end{figure}

\begin{figure}[h]
\centering
\includegraphics[angle=-90,scale=0.3]{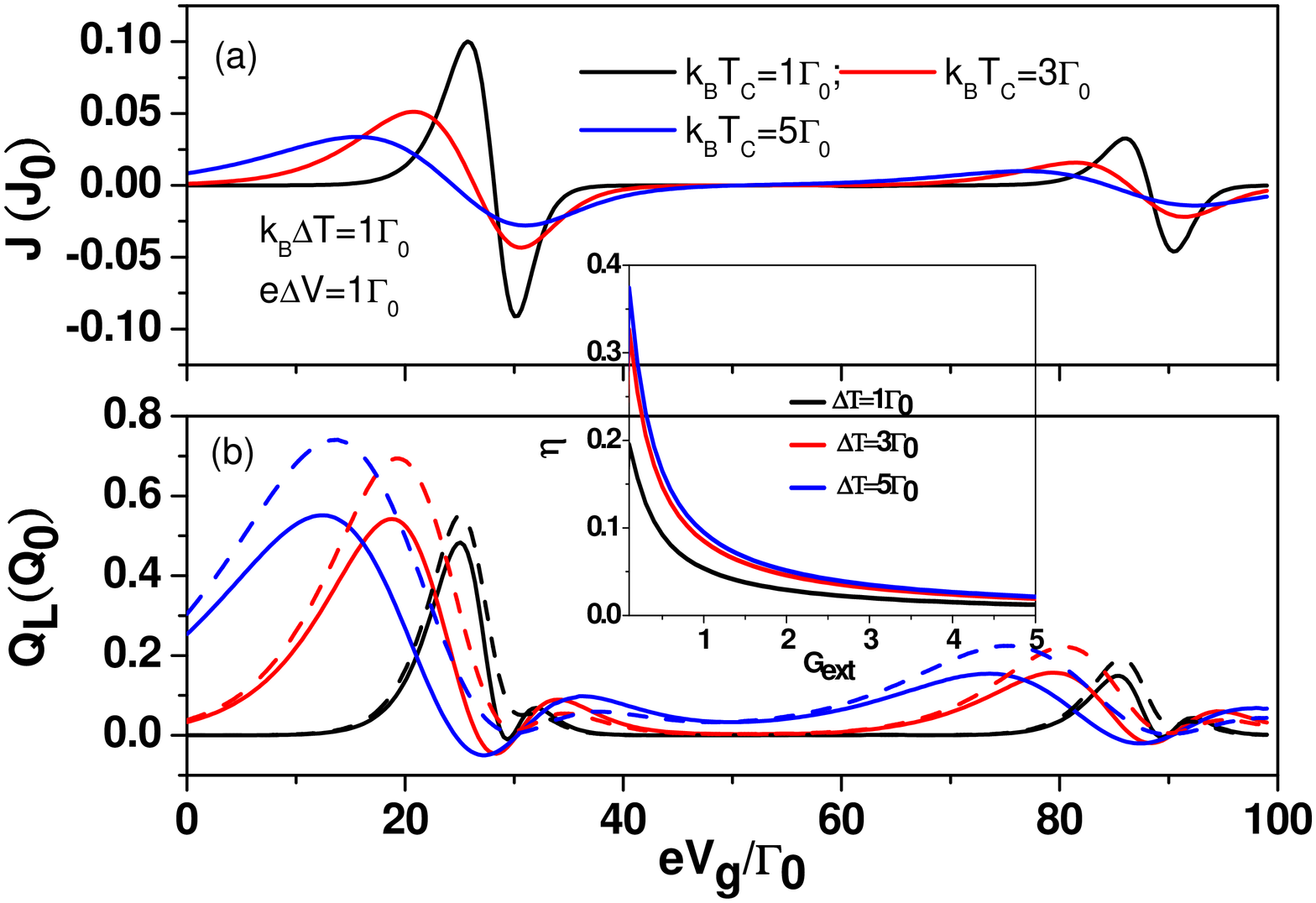}
\caption{Results obtained by method B for (a){electron current}
($J$), and (b) heat current ($Q_L$ in units of $Q_0=\Gamma_0^2/h$)
of SCTQD as functions of gate voltage $V_g$
($E_{\ell}=E_F+30\Gamma_0-eV_g$) for various values of $T_C$ with
$k_B\Delta T=1\Gamma_0$. Solid curves are for $\Delta V$ fixed at
$1 \Gamma_0$, while the dashed curves in (b) are for a fixed load
with $G_{ext}=0.2G_0$. Other physical parameters are the same as
those used for Fig. 4. The inset shows the $\eta$ as functions of
$G_{ext}$ for different $k_B\Delta T$ values at $eV_g=28\Gamma_0$
and $k_BT_C=1\Gamma_0$.}
\end{figure}

\begin{figure}[h]
\centering
\includegraphics[angle=-90,scale=0.3]{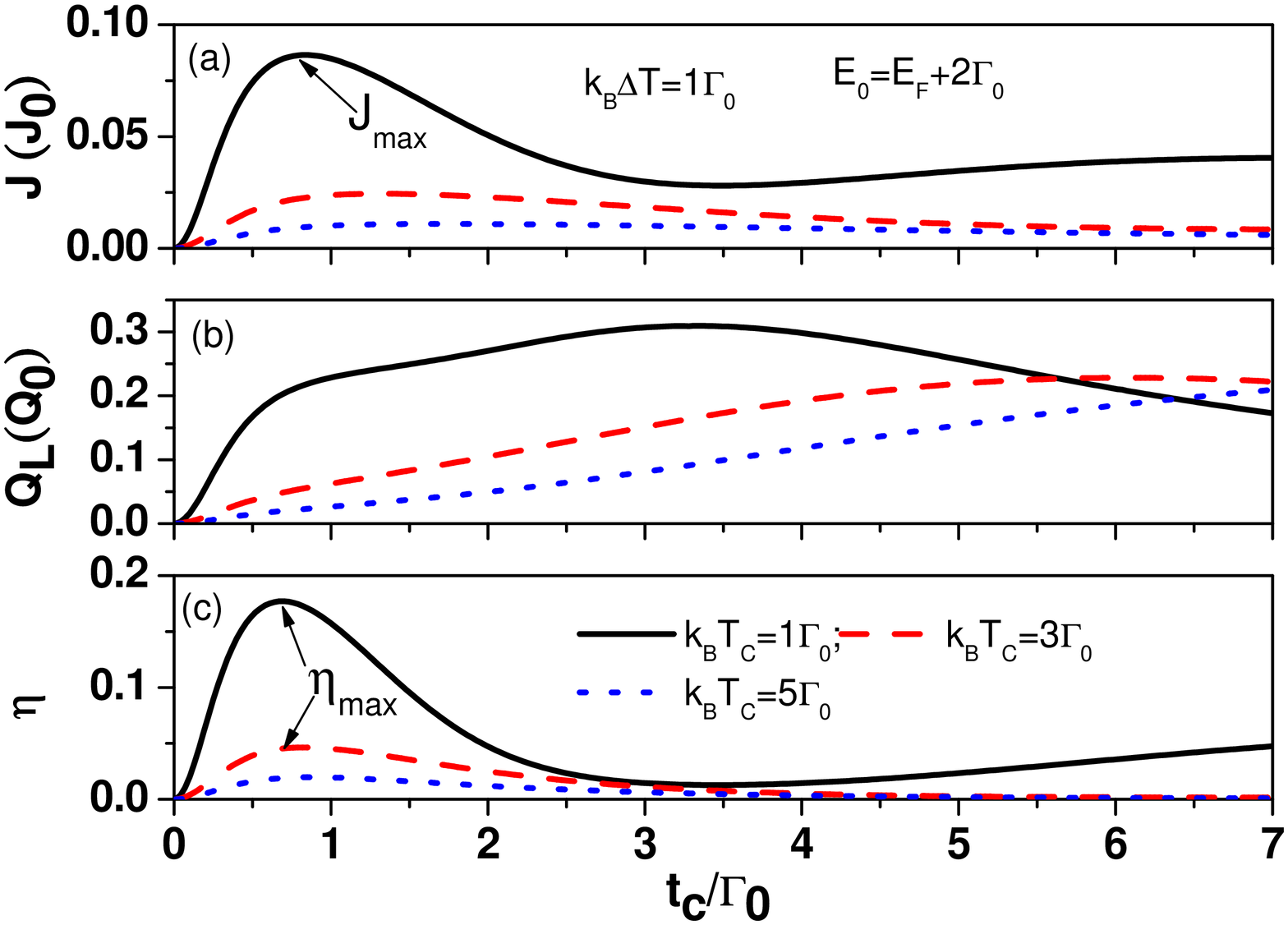}
\caption{Results obtained by method B for (a){electron current}
($J$),  (b) heat current ($Q_L$ in units of $Q_0=\Gamma_0^2/h$),
and (c) EHE efficiency of SCTQD as functions of electron hopping
strength $t_{LC}=t_{CR}=t_C$  for various values of $T_C$ with
$k_B\Delta T=1\Gamma_0$. Other physical parameters are the same as
those used for Fig.5.  To find the ratio of $\eta$ to the Carnot
efficiency, we should multipy $\eta$ in (c) by a scaling factor
$(T_C+\Delta T)/\Delta T$, which is 2, 4, and 6 for $k_B T_C=1,
3$, and  $5 \Gamma_0$,respectively.  }
\end{figure}

\begin{figure}[h]
\centering
\includegraphics[angle=-90,scale=0.3]{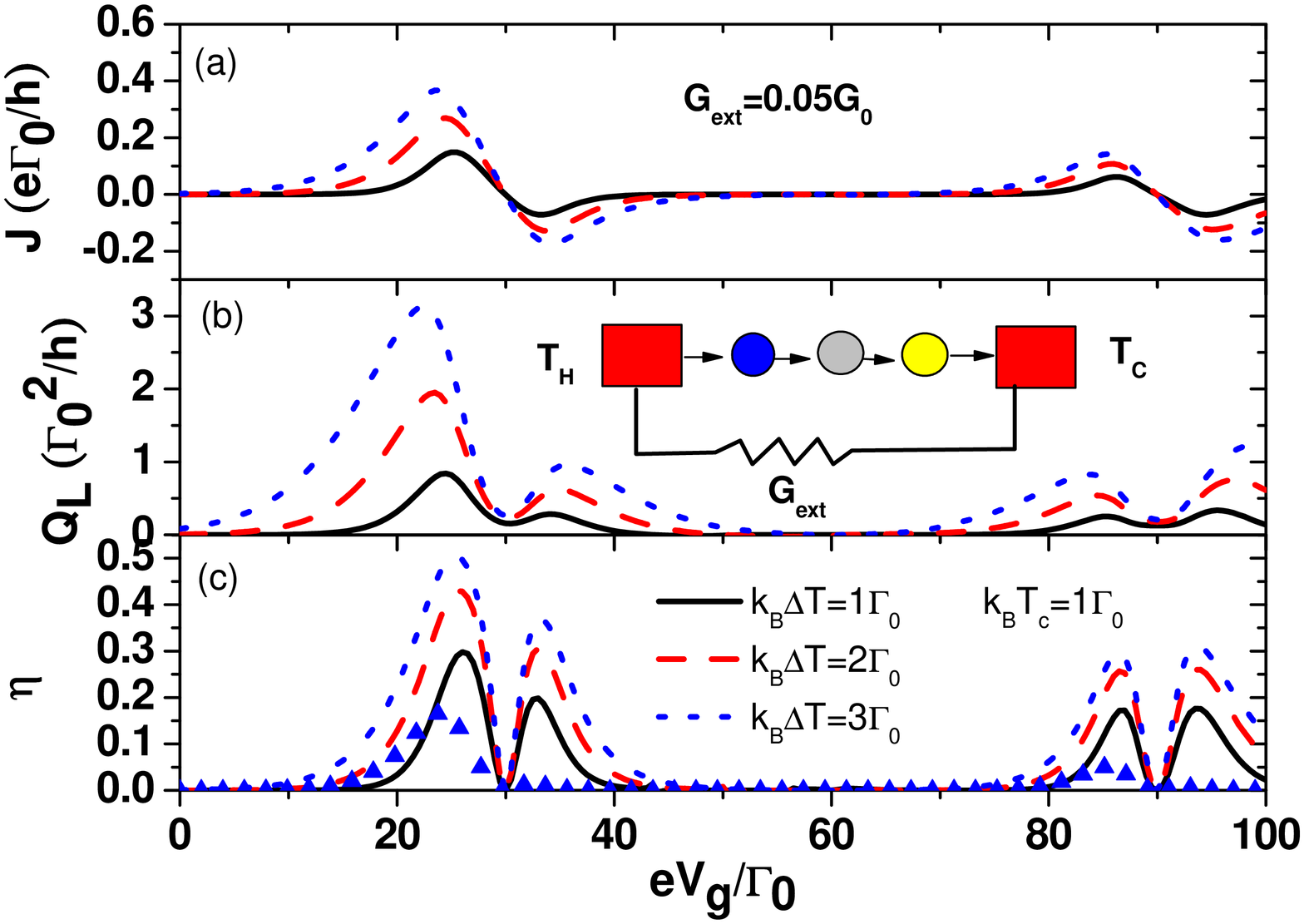}
\caption{(a) {Electron} current ($J$), (b) heat current ($Q_L$)
and (c) EHE efficiency $\eta$ as functions of QD energy level for
different $k_B\Delta T$ values at $k_BT_C=1\Gamma_0$ and
$G_{ext}=0.05G_0$. Other physical parameters are the same as those
of Fig. 3. The curve with blue triangle marks shown in Fig. 7(c)
is calculated by method B for the case of $k_B\Delta T=3\Gamma_0$
to reveal the $Q_{ph}$ effect. To find the ratio of $\eta$ to the
Carnot efficiency, we should multipy $\eta$ in (c) by a scaling
factor $(T_C+\Delta T)/\Delta T$, which is 2, 3, and 4 for $k_B
T_C=1, 2$, and  $3 \Gamma_0$,respectively.}
\end{figure}

\begin{figure}[h]
\centering
\includegraphics[angle=-90,scale=0.3]{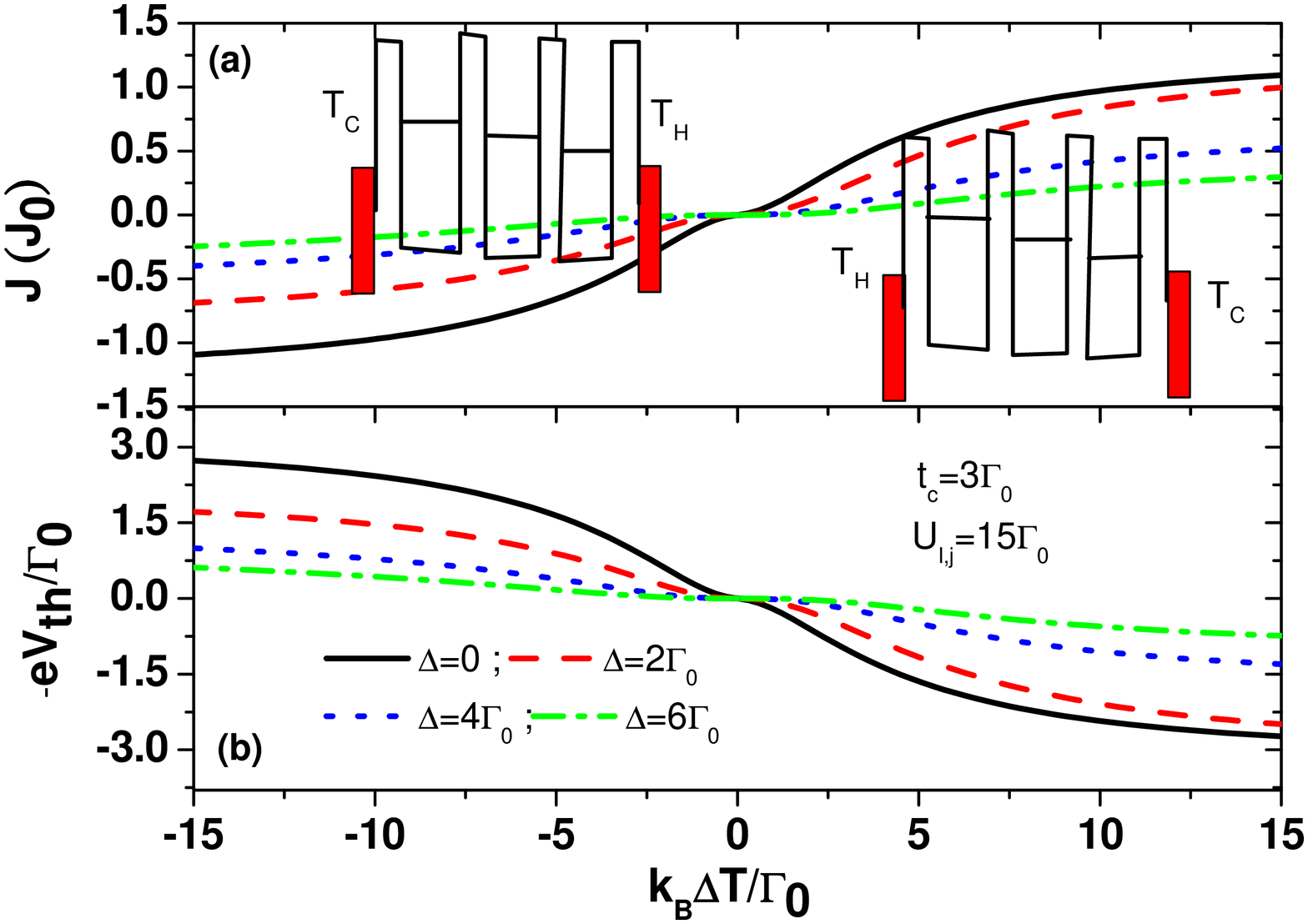}
\caption{(a) {Electron} current ($J$) and (b) thermal voltage
($eV_{th}$) as functions of temperature bias for different QDM
configurations ($E_R=E_F+10\Gamma_0$, $E_C=E_R+\Delta$, and
$E_L=E_R+2\Delta$) with $t_c=3\Gamma_0$, $U_{\ell,j}=15\Gamma_0$,
and {$k_BT_c=1\Gamma_0$}. We have considered QD energy levels
shifted by $V_{th}$. Here $E_{L(R)}$ is replaced by
{$\epsilon_{L(R)}=E_{L(R)}\mp 0.3eV_{th}$}. Other physical
parameters are the same as those of Fig. 3.  }
\end{figure}

\begin{figure}[h]
\centering
\includegraphics[angle=-90,scale=0.3]{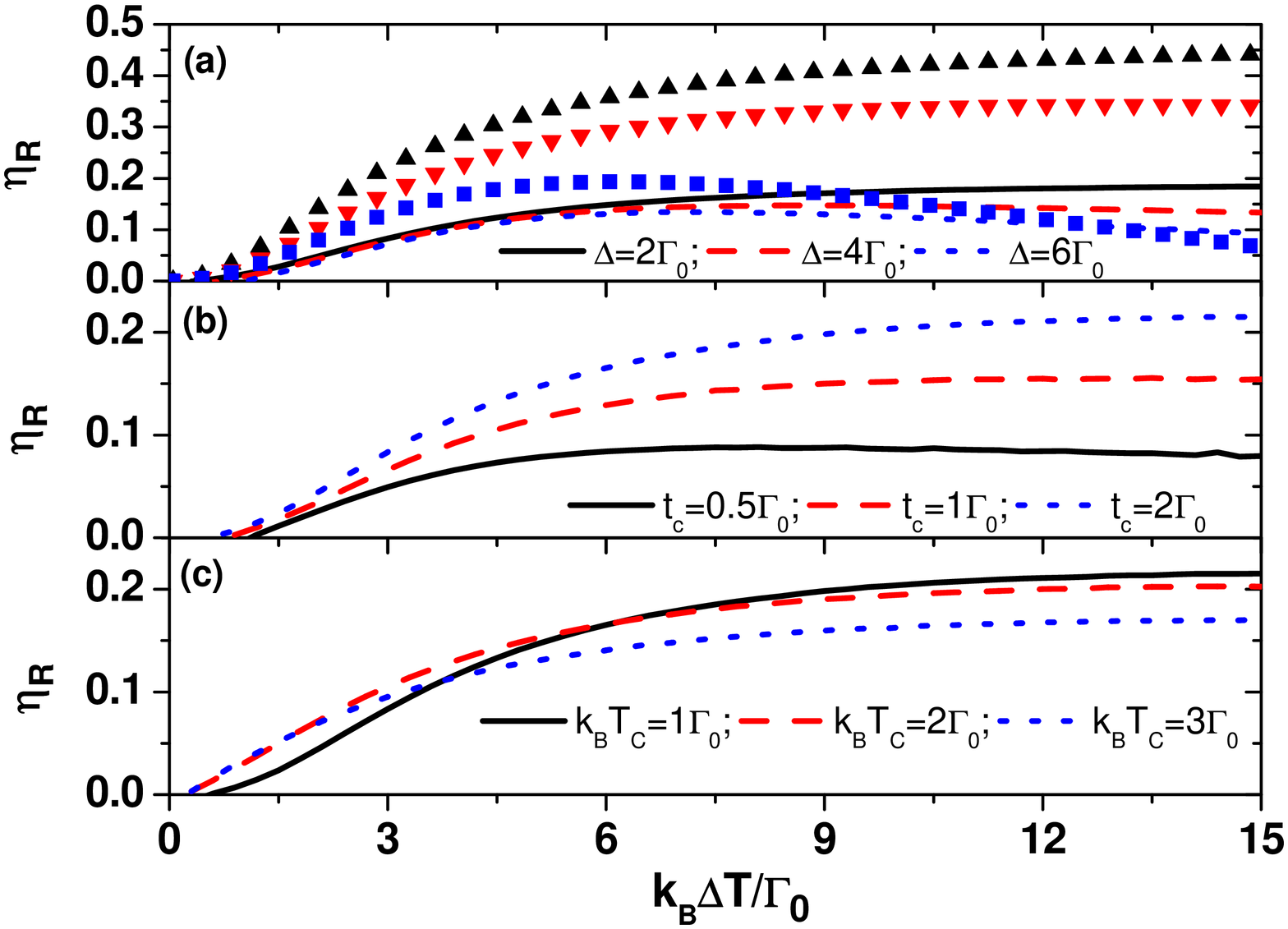}
\caption{ Electron current rectification efficiency ($\eta_R$)  as
a function of temperature bias for the variation of different
physical parameters. (a) $\Delta $ is varied, while
$t_c=3\Gamma_0$ and $k_BT_c=1\Gamma_0$. (b) $t_c$ is varied, while
$\Delta=2\Gamma_0$ and $k_BT_c=1\Gamma_0$.  (c) $T_c$ is varied,
while $\Delta=2\Gamma_0$ and $t_c=2\Gamma_0$. Other physical
parameters are the same as those of Fig. 8.}
\end{figure}

\clearpage

\end{document}